# Stochastic processes in light-assisted nanoparticle formation

Makoto Naruse[1,*], Yang Liu[2], Wataru Nomura[2], Takashi Yatsui[2],

Masaki Aida[3], Laszlo B. Kish[4], and Motoichi Ohtsu[2]

1 *Photonic Network Research Institute, National Institute of Information and Communications Technology, 4-2-1 Nukui-kita, Koganei, Tokyo 184-8795, Japan*

2 *Department of Electrical Engineering and Information Systems and Nanophotonics Research Center, Graduate School of Engineering, The University of Tokyo, 2-11-16 Yayoi, Bunkyo-ku, Tokyo 113-8656, Japan*

3 *Tokyo Metropolitan University, 6-6 Asahigaoka, Hino, Tokyo 191-0065, Japan*

4 *Department of Electrical and Computer Engineering, Texas A&M University, College Station, TX 77843-3128, USA*

**Abstract:** Recently, light-assisted nanofabrication have been introduced, such as the synthesis of quantum dots using photo-induced desorption that yields reduced size fluctuations, or metal sputtering under light illumination resulting in self-organized, nanoparticle chains. The physical mechanisms have originally been attributed to material desorption or plasmon resonance effects. However, significant stochastic phenomena are also present that have not been explained yet. We introduce stochastic models taking account of the light-assisted processes that reproduce phenomenological characteristics consistent with the experimental observations.

Nanophotonic devices and systems have been intensively studied for use in a wide range of applications, such as information and communication, energy and the environment, etc.[1,2] Precision control of the geometrical features of materials on the nanometer scale, such as the sizes and positions, are important factors in obtaining the intended functionalities of nanophotonic devices and systems in which multiple nanostructures are mediated by optical near-field interactions,[1] and also for plasmonic devices.[2] To satisfy such requirements, light-assisted, self-organized nanostructure fabrication principles and techniques have been developed.[3,4] One example is the sol-gel synthesis of ZnO quantum dots (QDs) using photo-induced desorption, which yields reduced QD diameter fluctuations.[3] (In Ref. 3, they are called "variations". We use the term "fluctuations" throughout the paper with the same meaning.) From an application standpoint, the sizes of QDs should be well-controlled to ensure that the quantized energy levels are resonant between adjacent QDs, facilitating efficient optical near-field interactions.[5] Another example of light-assisted nanostructure fabrication is metal sputtering with light irradiation, which yields self-organized, size- and position-controlled metal nanoparticle chains.[4] Arrays of nanoparticles are important in various applications, such as nanophotonic devices,[1] optical far-field to near-field converters,[6] plasmonic light transmission lines,[7] etc. We should also emphasize that common advantages of these light-assisted, self-organized fabrication techniques are their relatively simple experimental setups and superior production throughput compared with, for instance, scanning-based methods, such as those based on electron beams[8] or scanning probes.[9]

The physical mechanisms behind the light-assisted nanostructure formation have been attributed to material desorption[3,10,11] or plasmon resonance between light and matter.[4] However, stochastic physical processes are also involved, as observed in the experimental data reported below. Also, we consider that stochastic approaches are indispensable to take account of the emergence of ordered structures and the wide range of phenomena observed on the nanoscale in general.[12,13] For example, Söderlund *et al.* demonstrated lognormal size distributions in particle growth processes with a simple statistical model,[12] and Kish *et al.* demonstrated the lognormal distribution of single-molecule fluorescence bursts in micro- and nano-fluidic channels based on a stochastic analysis.[13] Also, a study[14] of the stochastically driven growth of self-organized structures indicates that the spatio-temporal distribution functions have a key role in controlling the shape and width of size distributions within the formations. Cutting the log tails of such distribution functions can contribute to narrower size distributions.

In this Letter, we approach light-assisted nanofabrication from a stochastic standpoint. We build stochastic models taking account of the light-assisted processes that reproduce tendencies consistent with experimental observations. Through such considerations, we obtain critical insights into the order formation on the nanometer scale, which will contribute to the design of nanophotonic devices and systems. Before the discussion, note that the term "size" is used when it is relevant to general indications of physical size, including diameter, whereas the term "diameter" is used when it is relevant to specific experimental results discussed in this Letter.

Firstly, we characterize the light-assisted, self-organized ZnO quantum dot formation, which was experimentally demonstrated in Ref. 3, with a stochastic approach. We first briefly review the experimental observations that have been reported.

Among various methods of fabricating ZnO QDs, synthetic methods using liquid solutions are advantageous because of their need for simple facilities and their high productivity[15] compared with those based on laser ablation,[16] reactive electron beam evaporation,[17] etc. Although the size of the QDs, which is precisely the diameter of the QDs, fluctuates by as much as 25% in conventional sol-gel methods,[15] the light-assisted sol-gel method demonstrated in Ref. 3 reduced the QD diameter fluctuations. When light with a photon energy higher than the bandgap energy is radiated during the ZnO QD formation process, electron–hole pairs could trigger an oxidation–reduction reaction in the QDs, causing the ZnO atoms depositing on the QD surface to be desorbed. In addition, such desorption is induced with a high probability when the formed QDs reach a particular diameter. This light-dependent QD size regulation has also been reported in other material systems, such as CdSe[10] and Si.[11]

The insets in Fig. 1 (a) and (b) respectively show transmission electron microscope (TEM) images of fabricated ZnO QDs without and with continuous-wave (CW) light illumination at a wavelength of 325 nm with a power density of 8 mW cm$^{-3}$. The experiments are detailed in Ref. 5. Figure 1 (a) and (b) respectively summarize the incidence rate of nanoparticles as a function of their diameter. The diameter fluctuations decreased from 23% to 18% with light irradiation.

What we particularly address in this Letter is that the diameter distributions are different between these two cases. It exhibits behavior similar to a normal distribution without light illumination (Fig. 1(a)), whereas the distribution is skewed with light irradiation; in particular, the incidences at larger diameters decreased (Fig. 1(b)). We investigate the different behavior by means of stochastic modeling, as described below.

First, in the absence of light illumination, we represent the formation process as a statistical pile-up model, as schematically shown in Fig. 2(a). An elemental material that constitutes a nanoparticle is represented by a square-shaped block. Such blocks are grown, or stacked one after another, with a piling success probability $p$; accordingly, the piling fails with a probability of $1-p$. In other word, if we let the length of the pile at step $t$ be $s(t)$, the piling probability is given by

$$P[s(t+1) = s(t)+1 | s(t)] = p$$
$$P[s(t+1) = s(t) | s(t)] = 1-p. \qquad (1)$$

Since this is equivalent to a random walk with drift, after repeating this process with an initial condition $s(0) = 0$, the resultant lengths of the piles exhibit a normal distribution, as shown in Fig. 1(c). Specifically, the statistics shown in Fig. 1(c) were obtained by repeating 10,000 steps for 100,000 different trials.

On the other hand, we model the effect of light irradiation in the formation process in the stochastic model as follows. As described above, since the material desorption is likely induced at a particular diameter of nanoparticle,[5] we consider that the piling success rate $p$ is a function of the diameter, namely the height of the pile. For simplicity, we consider that $p$, which represents the deposition success probability, decreases linearly beyond a certain total height of a pile, as schematically shown in Fig. 2(b). In other words, the material desorption is more likely to be induced beyond a certain pile size due to the resonant effect mentioned above. That is, the probability $p$ in Eq. (1) is replaced with the following size-dependent probability;

$$p(s(t)) = \begin{cases} c & s(t) \leq R \\ c - \alpha s(t) & s(t) > R \end{cases} \qquad (2)$$

where $c$ and $\alpha$ are constants. With such a stochastic model, the resultant incidence distributions of the piles is skewed or reduced at larger sizes. In the calculated results shown in Fig. 1(d), we assume $c = 1/2$ and $\alpha = 1/250$. The numerical results obtained through the statistical modeling are consistent with the experimental observations.

Secondly, in Ref. 4, self-organized formation of an array of ultralong nanoparticle chains was demonstrated based on near-field optical desorption. We first briefly describe our experimental observations. With conventional radio-frequency (RF) sputtering, we deposited aluminum on a glass substrate. A 100 nm-wide and 30 nm-deep groove was formed in the substrate, as schematically shown in Fig. 3(a). Also, the substrate was illuminated with light linearly polarized perpendicularly to the direction of the groove during the RF sputtering. Thanks to the edge of the groove, a strong optical near-field was generated in its vicinity.

A metallic nanoparticle has strong optical absorption because of plasmon resonance[18–20], which depends strongly on the particle size. This can induce desorption of a deposited metallic material when it reaches the resonant diameter.[21,22] It turns out that as the deposition of the metallic material proceeds, the growth is governed by a tradeoff between deposition and desorption, which determines the particle diameter, depending on the photon energy of the incident light. Consequently, an array of metallic nanoparticles is aligned along the groove, as shown in Fig. 3(b). While radiating continuous-wave (CW) light with a photon energy of 2.33 eV (wavelength: 532 nm) during the deposition of aluminum, 99.6 nm-diameter, 27.9 nm-separation nanoparticles were formed in a region as long as 100 μm, as shown in Fig. 3(b).

As described above, the origin of the size regulation of the nanoparticles was attributed to the resonance between the nanoparticles and the illuminated light, similarly to the case discussed earlier. At the same time, we consider that although this physical mechanism indeed plays a crucial role, it is not enough to explain the formation of the uniformly formed array structure. To explain such an observation, we need to extend the stochastic model described above as follows.

In the modeling, we assume a one-dimensional horizontal system that mimics the groove on the substrate. More specifically, it consists of an array of *N* pixels with their identity represented by an index *i* ranging from 1 to *N*. An elemental material to be deposited onto the system, experimentally by the RF sputtering described above, is schematically represented by a square-shaped block. As depicted in Fig. 4(a), the initial condition is a flat structure without any blocks.

At every iteration cycle, the position at which a block arrives is randomly chosen, which we denote *x*. We determine the success of the deposition at *x* by the following rules. We denote the occupation by a block at position *x* of the groove by $S(x)$: $S(x)=1$ when a block occupies a position *x,* and $S(x)=0$ when there is no block at position *x*. Also, we use the term "cluster" to mean multiple blocks consecutively located along the groove. We also call a single, isolated block in the system a "cluster".

(i) When the randomly chosen position *x* belongs to one of the cluster(s), namely, $S(x)=1$, we maintain $S(x)=1$. (Fig. 4(b,i))

(ii) Even if $S(x)=0$, when the chosen position *x* belongs to a "neighbor" of a cluster with a size greater than a particular number $B_{th1}$, the deposition is inhibited. That is, we maintain $S(x)=0$. (Deposition is inhibited.) (Fig. 4(b,ii))

(iii) Even if $S(x)=0$, when the chosen position *x* has blocks at both its left and right sides and the total number of connected blocks is greater than $B_{th2}$, the deposition is inhibited. That is, we maintain $S(x)=0$. (Fig. 4(b,iii))

(iv) In other cases, the deposition at the position *x* succeeds; namely, $S(x)=1$. (Fig. 4(b,iv))

The rules (ii) and (iii) correspond to the physical effect of the resonance between the material and the light illumination that facilitates desorption of the particle. The optical near-field intensity in the vicinity of a nanostructure follows a Yukawa function[3] which depends on the material size. Therefore, the optical near-fields promote material desorption, or in effect, inhibits material deposition, beyond a certain size of nanoparticles, which is characterized as rule (ii) above. Also, even when a single cluster size is small, meaning that the corresponding near-fields are small, when several such clusters are located in close proximity, a material desorption effect should be induced overall. Such an effect is represented as rule (iii) above. One remark here is that we do not pile more than two blocks at a single position $x$; that is to say, $S(x)$ takes binary values only, since our concern is how the clusters are formed in the 1D system.

Figure 5 shows a numerical demonstration assuming a 1D array with $N = 1000$. As statistical values in the simulations, we evaluated the incidence of the cluster size and the center-to-center interval between two neighboring clusters. Figure 5 (a) and (b) summarize the evolution of these two values at $t = 100$, $t = 1,000$, $t = 10,000$, and $t = 100,000$. In the numerical calculations, for the threshold values in rules (ii) and (iii), we assumed $B_{th1} = 8$ and $B_{th2} = 12$, respectively. We clearly observed that the size and the interval converged to representative values, which are consistent with the experimental observations shown in Fig. 3(b).

Furthermore, as reported in Ref. 4, a similar experiment was conducted with a higher photon energy of 2.62 eV (473 nm) an an optical power of 100 mW, which yielded a 84.2 nm-diameter, 48.6 nm-separation nanoparticle formation. As summarized in Table 1, the diameter is slightly reduced and the nanoparticle distance is enlarged compared with the previous case of lower photon energy (2.33 eV (532 nm)). The reduced diameter of the nanoparticles is attributed to the fact that the higher photon energy leads to desorption at smaller diameters.[4,18] The larger separation between adjacent nanoparticles is, however, not obviously explained.

We presume that a stronger light–matter resonance is induced at a higher photon energy, which more strongly induces material desorption, or inhibits the deposition of materials, in the neighboring clusters. We can take account of this effect by modifying the stochastic model described above. Instead of blocking the deposition at the neighboring positions by rule (ii), we consider that distant neighbors are also inhibited:

(ii') Even if $S(x) = 0$, when $x$ sees a cluster with a size greater than a particular number $B_{th1}$, within a area (a) between $x-3$ and $x-1$ or (b) between $x+1$ and $x+3$, the deposition is inhibited. That is, we maintain $S(x) = 0$.

While preserving $B_{th1}$ and $B_{th2}$ values with the previous example, the cluster size statistics evolve as shown in Fig. 5 (c). At the iteration cycles $t$ = 10,000 and 100,000, the incidences of single-sized clusters are large. This is due to the strict inhibition rule (ii') above, which reduces the chance of clusters growing. Treating such a single-sized cluster as an artifact, or a virtually ignorable element, in the system, we evaluate the cluster-to-cluster interval except for single-sized clusters. The cluster interval converges to a maximum of 10, as shown in Fig. 5(d), which is larger than the previous case which converged to 8, as shown in Fig. 5(a). This is consistent with the experimental observations. Finally, we make one remark about the dimensions of the models concerned in this Letter. We consider that the 1D models described above characterize the physical principles behind the experimental demonstrations of ZnO QD formation and the Al nanoparticle array formation.

In summary, we developed stochastic models taking account of the optical near-field-based material desorption/deposition between light and matter on the nanometer scale. By using a simple model, the observed behavior of skewed diameter distributions of ZnO quantum dots with light irradiation, and self-organized nanoparticle array formation along a groove were reproduced in the stochastic modeling, providing greater insights into the order formation occurring on the nanometer scale. We consider that such modeling will be applicable to other light-assisted material formation processes, and will be beneficial in the design of future nanophotonic devices and systems. For example, phonon-assisted near-field processes generated characteristic surface morphology [23]. Further investigation based on a stochastic viewpoint will be more important in the future.

This work was supported in part by Grants-in-Aid for Scientific Research from the Japan Society for the Promotion of Science (JSPS) and the Strategic Information and Communications R&D Promotion Programme (SCOPE) of the Ministry of Internal Affairs and Communications, Japan.

**TABLE 1.** The diameter and the interval of the nanoparticles obtained with light-assisted Al sputtering.

| Assist light | Diameter (nm) | Nanoparticle interval (nm) |
|---|---|---|
| 2.33 eV (532 nm) | 99.6 | 127.5 |
| 2.62 eV (473 nm) | 84.2 | 132.8 |

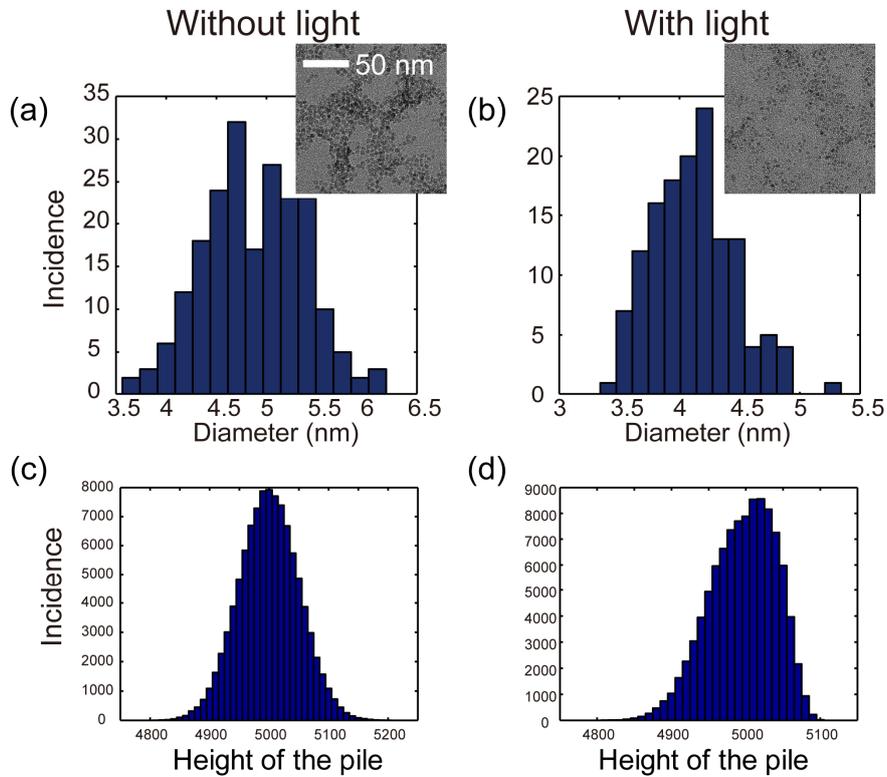

**FIG. 1.** (Color online) (a,b) Incidence patterns of the diameters of fabricated ZnO quantum dots (QDs) formed by a sol-gel method (a) without light illumination and (b) with light irradiation. Insets in (a) and (b) respectively are transmission electron microscope images of QDs without and with light illumination. With light irradiation, the incidences of the larger-diameter QDs are reduced, i.e., the diameter distribution is skewed. (c,d) Incidence patterns of the size distribution generated with the proposed stochastic models. The patterns are consistent with the experimental observations in (a) and (b).

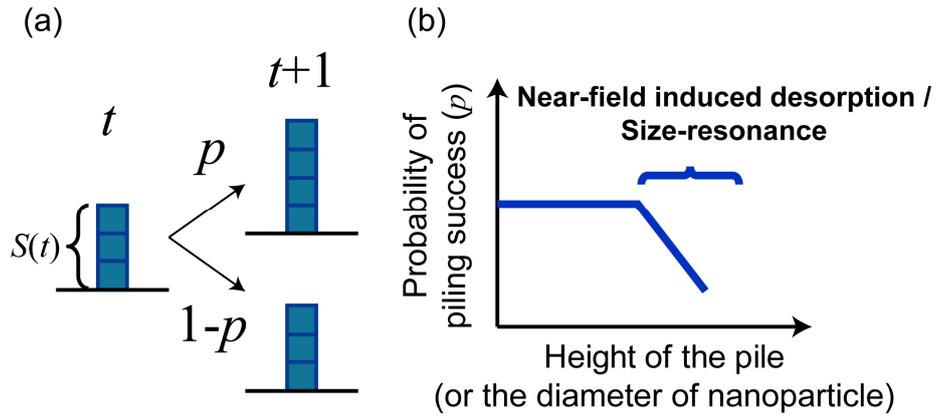

**FIG. 2.** (Color online) A stochastic model of light-assisted nanoparticle formation. (a) The growth of the QD is characterized with a one-dimensional pile-up model. The success of the piling depends on probability *p*. (b) The effect of light irradiation is modeled by a decrease in the probability *p* beyond a certain size of pile, which corresponds to the diameter in the experiment.

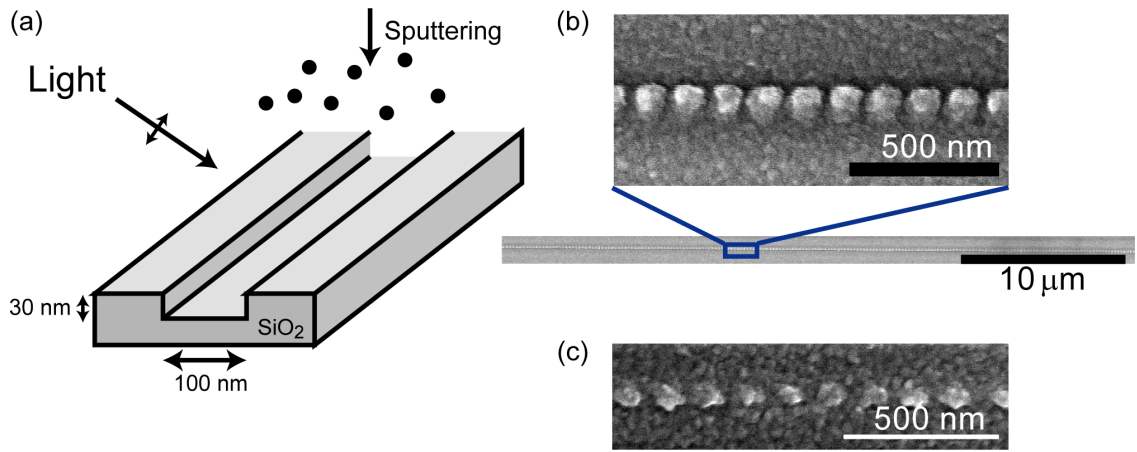

**FIG. 3.** (Color online) (a) Schematic diagram of the experimental setup of Al sputtering on an SiO$_2$ substrate in which a 100 nm-wide, 30 nm-deep groove is formed. During the sputtering, the substrate is irradiated with light having a polarization perpendicular to the direction of the groove. (b,c) An array of uniform-diameter, uniform-separation Al nanoparticles is self-organized along the groove, with (b) 2.33 eV light irradiation and (c) 2.62 eV light irradiation.

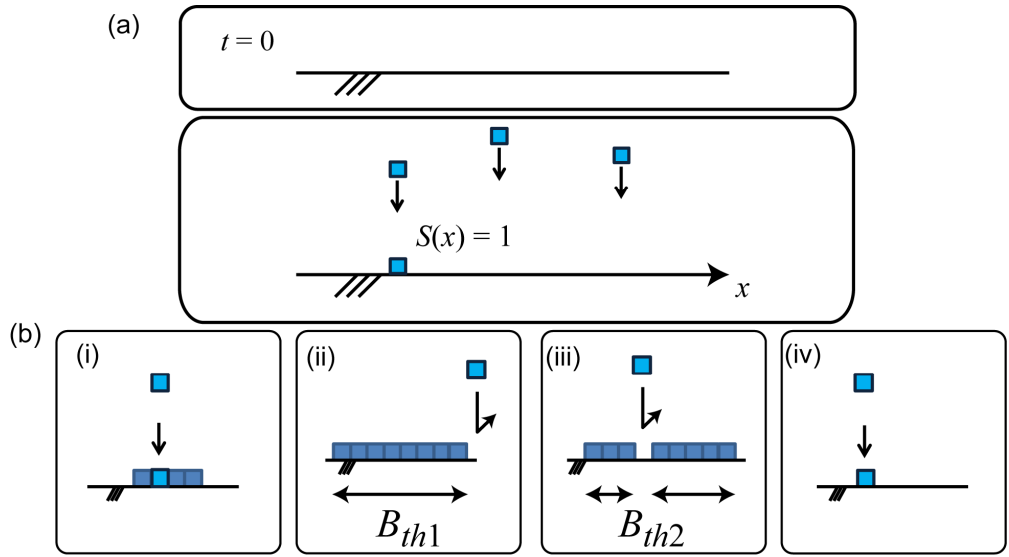

**FIG. 4.** (Color online) A stochastic model of the nanoparticle array formation. (a) One-dimensional array in which an elemental block could be deposited at position $x$. (b) Rules for successful deposition at a randomly chosen position $x$. (b,ii) Deposition is inhibited in neighboring clusters whose size is larger than $B_{th1}$. (b,iii) Deposition is inhibited at positions where they see clusters at both the left- and right-hand sides and when the total size of both clusters is larger than $B_{th2}$.

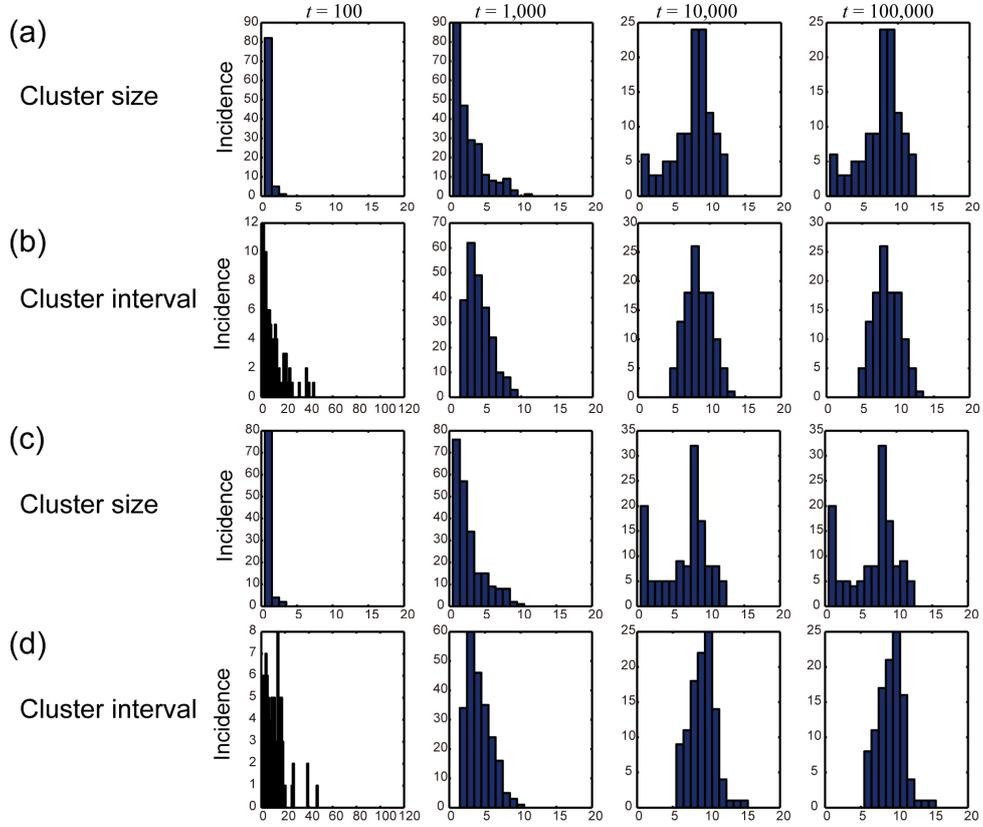

**FIG. 5.** (Color online) Evolution of (a) the cluster size and (b) the cluster interval based on a stochastic model. Both the size and the interval converge to incidence patterns that exhibit their maximum at a particular value, which reproduced the size- and separation-controlled, nanoparticle array formations observed experimentally. (c,d) Note that the separation of the nanoparticles is greater with higher photon energy (Fig. 3(c), Table 1). By modifying rule (ii) of the stochastic modeling, the cluster interval increases, which is consistent with the experimental observations.